\documentclass[reprint, amsmath,amssymb,aps,superscriptaddress]{revtex4-2}
\pdfoutput=1
\usepackage{graphicx}
\usepackage{subcaption}
\usepackage{caption,color}
\usepackage{bbold}
\usepackage{cleveref}
\usepackage{braket}
\usepackage{dcolumn}
\usepackage{bm}
\usepackage{breqn}
\usepackage[font={small,it}]{caption}

\def\<{\langle}
\def\>{\rangle}

\newcommand{\be}{\begin{equation}}
\newcommand{\ee}{\end{equation}}
\newcommand{\ba}{\begin{array}}
\newcommand{\ea}{\end{array}}
\newcommand{\bqa}{\begin{eqnarray}}
\newcommand{\eqa}{\end{eqnarray}}

\newcommand{\Imperial}{Blackett Laboratory, Imperial College London, SW7 2AZ, United Kingdom}
\newcommand{\HZDR}{Helmholtz-Zentrum Dresden-Rossendorf, Bautzner Landstraße 400, 01328 Dresden, Germany}
\newcommand{\sussex}{Sussex Centre for Quantum Technologies, University of Sussex, Brighton, BN1 9RH, United Kingdom}
\newcommand{\bristol}{Quantum Engineering Centre for Doctoral Training, University of Bristol, Bristol, BS8 1TH, United Kingdom}

\begin{document}

\title{Optimal control with a multidimensional quantum invariant}
\author{Modesto Orozco-Ruiz}\affiliation{\Imperial}
\author{Selwyn Simsek}\affiliation{\Imperial}
\author{Sahra A. Kulmiya}\affiliation{\sussex}\affiliation{\bristol}
\author{Samuel J. Hile}\affiliation{\sussex}
\author{Winfried K. Hensinger}\affiliation{\sussex}
\author{Florian Mintert}\affiliation{\Imperial}\affiliation{\HZDR}

\date{\today}
\begin{abstract}
Optimal quantum control of continuous variable systems poses a formidable computational challenge because of the high-dimensional character of the system dynamics.
The framework of quantum invariants can significantly reduce the complexity of such problems, but it requires the knowledge of an invariant compatible with the Hamiltonian of the system in question. We explore the potential of a Gaussian invariant that is suitable for quadratic Hamiltonians with any given number of motional degrees of freedom for quantum optimal control problems that are inspired by current challenges in ground-state-to-ground-state shuttling of trapped-ions.
\end{abstract}

\maketitle

\section{Introduction} \label{Introduction}

Quantum optimal control is widely accepted as one of the central tools in the development of quantum technological applications~\cite{koch22, wers07}.
While control of discrete degrees of freedom, such as qubits, is well established both in theory and experiment~\cite{bruz19},
the control over motional degrees of freedom is still in an early stage of development~\cite{bruz19, kiel02}. In particular, the identification of optimal solutions for the control of motional quantum states is rather computationally challenging due to the strictly infinite-dimensional, or practically high-dimensional, Hilbert space of the underlying control problem. 

While most proof-of-principle demonstrations of elementary building blocks of quantum information processing did not require control over motional quantum states beyond a mechanism that holds qubits in place, the current challenges in developing scalable technologies lead to a growing need to control motional quantum states \cite{ladd10}.
In trapped-ion hardware, for example, scalability is expected to require modular architectures~\cite{leki16,roos99,hens05} that allow the ions, which act as qubits, to be separated into many small groups.
Interconnecting spatially separated clusters with quantum logical operations then requires the ability to move individual ions from one cluster to another \cite{hucu08}.
Since trapped-ion quantum logic requires the motional states of the ions to be of sufficiently low energy, it is essential that such shuttling operations do not result in motional excitations at the end of the process.
This can be achieved with adiabatically slow shuttling, but, in practice, diabatic protocols~\cite{hens05,leki16,monr13} are sought-after because of the requirement to realize all operations of a quantum algorithm within the system's coherence time. 


Transport along a line can be performed diabatically,
and diabatic ground-state to ground-state transport is within experimental capabilities~\cite{bowl12, walt12}.
Diabatic transport beyond such one-dimensional problems, has been experimentally demonstrated in favorable geometries~\cite{kauf17}.
In typical trapping geometries, however, actual realizations of diabatic transport protocols result in excess motional energy in the final state.

Many diabatic protocols that ensure that the controlled object ends up in its quantum mechanical ground state are found within the concept of \textit{shortcuts to adiabaticity}~\cite{guery19} (STA). Among the different STA strategies, the Ermakov-Lewis quantum dynamical invariants~\cite{erm80,lewi68,lewi69,khan79} are particularly useful as they provide a framework for inverse-engineering appropriate control Hamiltonians~\cite{chen10,guery19,chen11,chen11-2, muga10,xiao18,palm15,palm15-2, palm16,selw21,selw21-2, furs14} as well as providing physical insight into the quantum system dynamics. 

Quantum invariants have been applied to a wide variety of theoretical control problems, from atom cooling~\cite{chen10}, fast separation of two trapped-ions~\cite{palm15-2}, uni-dimensional atomic transport in harmonic-traps~\cite{chen11-2, furs14} or expansions and compressions of trapped-ion chains with minimal motional excitation~\cite{palm15}. Combined with variational methods or included in the master equation formalism, they have also been shown to be useful in addressing problems that do not admit a standard treatment, such as anharmonic-potential transport~\cite{li22}, experimental noise minimization~\cite{levy18} or atomic state population inversion~\cite{whit20}.

While a substantial part of the conceptual development of quantum invariants is devoted to the construction of invariants for systems with only one translational degree of freedom, many practical applications of this framework require invariants for systems with more translational degrees of freedom. Here, we focus on a recently developed Gaussian invariant for multi-dimensional systems~\cite{selw21}, and explore its suitability for applications in quantum optimal control with theoretical examples inspired by the current ion shuttling experiments. 

\section{Invariant-based Inverse Engineering Method} \label{Invariant}

This section provides a short review over the invariant~\cite{selw21} used in the explicit examples discussed below in Sec.~\ref{results}.
In particular, we sketch how a time-dependent Hamiltonian can be constructed such that it induces dynamics with desired properties.

\subsection{Hamiltonian model}

Real-world potentials are never strictly harmonic, but if an object's quantum state is localized within a potential well, such as a sufficiently cooled ion, even an ostensibly anharmonic potential landscape can be well approximated as harmonic around the expected position of the ion. During the shuttling protocol, this quadratic approximation is valid as long as the localization in real-space remains sufficiently narrow.

The Hamiltonian for an ion of mass $m$, with a quantum state localized around a $d$-dimensional trajectory $\vec z(t)$, can be expressed as
\be \label{Hamiltonian_2}
\hat{H}(t) = \frac{\vec p^2}{2m} + \frac{1}{2}m\vec{x}^T\mathbf{M}\vec{x} - \vec{F}^T\vec{x}\ ,
\ee
where the upper-script $T$ represents the transpose operation, $\vec x$ and $\vec p\equiv -i\vec \nabla_x$ are the canonical displacement and momentum operators with respect to the trajectory $\vec z(t)$, and the vector $\vec F$ and matrix $\mathbf{M}$ correspond to the first and second derivatives of the trapping potential taken along $\vec z(t)$.

Up to an irrelevant scalar term, this quadratic Hamiltonian can also be cast as
\be
\hat{H}(t) = \frac{\vec p^2}{2m}+\frac{1}{2}m(\vec x-\vec{C}(t))^T\mathbf{M}(t)(\vec{x} - \vec{C}(t))\ ,
\label{Hamiltonian}
\ee
with the trap center $\vec C(t)=\frac{1}{m}\mathbf{M}^{-1}\vec F(t)$.

If the initial quantum state of the ion is Gaussian, such as the ground-state of the Hamiltonian with a harmonic trapping potential, or a thermal state,
the Gaussian character of the quantum state is preserved during the shuttling protocol within this quadratic approximation.
The dynamics are then characterized completely in terms of the classical trajectory $\vec z(t)$ and momentum $\vec p(t)$ (\textit{i.e.}, the expected value of the displacement and momentum operators $\vec{x}$ and $\vec{p}$) and the covariance matrix $\mathbf{\Sigma}(t)$ of all variances and covariances of $\vec x$ and $\vec p$.

\subsection{Multi-dimensional Gaussian Quantum Invariant}

The goal of shuttling an ion such that it ends up in the ground-state of its final Hamiltonian can be understood as an optimization problem,
with motional excitation as figure of merit.
Instead of finding a suitable protocol as result of an optimization, one can also design suitable shuttling protocols more directly with the framework of quantum invariants.

A quantum invariant $\hat{I}(t)$ is an operator that satisfies the equation of motion
\begin{eqnarray}
\label{invariant_eq}
\frac{\partial \hat{I}(t)}{\partial t} = i[\hat{I}(t), \hat{H}(t)]\ .
\end{eqnarray}
The property of such invariants that is crucial for quantum control is that the eigenstates of $\hat{I}(t)$ -- which are generally time-dependent -- are solutions of the time-dependent Schr\"odinger equation with the Hamiltonian $\hat{H}(t)$.

A non-degenerate eigenstate of an invariant $\hat{I}(t)$ that commutes with the Hamiltonian $\hat{H}(t)$ at the beginning and at the end of the dynamics thus corresponds to a shuttling protocol in which an ion starts in an eigenstate of the initial Hamiltonian $\hat{H}(t=0)$ and ends up in an eigenstate of the final Hamiltonian $\hat{H}(t=T)$ even though it does not need to follow the dynamics of an eigenstate as it would be the case in an adiabatic protocol.

A shuttling protocol can thus be defined in terms of a time-dependent invariant,
and Eq.~\eqref{invariant_eq} is the defining relation determining a time-dependent Hamiltonian $\hat{H}(t)$ that achieves the desired shuttling protocol.

The following analysis is based on an invariant~\cite{selw21,selw21-2} that is defined in terms of the actual classical trajectory $\vec{z}(t)$ that the ion is meant to take,
and a time-dependent, positive ({\it i.e.} positive semi-definite), $d$-dimensional matrix $\mathbf{R}(t)$.

The quadratic component $\mathbf{M}(t)$ of the potential is obtained from $\mathbf{R}(t)$ via the relation~\cite{selw21}
\bqa
\label{m_eq}
\{\mathbf{R}^2, \mathbf{M}\} = 2[\dot{\mathbf{R}}, \mathbf{R}]_{\mathbf{A}} - \{\ddot{\mathbf{R}}, \mathbf{R}\} - 2\mathbf{R}\mathbf{A}^2\mathbf{R}\ ,
\eqa
where $[X,Y]_Z=XZY-YZX$ is the generalized commutator, $\{X, Y \} = XY + YX$ the anticommutator and 
\bqa
\label{Aeq}
\mathbf{A}&=&i\mathbf{R}^{-2} + \frac{1}{2}[\mathbf{R}^{-1},\dot{\mathbf{R}}] + \frac{1}{2}\mathbf{R}^{-1}\mathbf{J}\mathbf{R}^{-1}\ ,\\
\label{Jeq}
\{\mathbf{J},\mathbf{R}^{-2}\} &=& [\dot{\mathbf{R}}, \mathbf{R}^{-1}] + [\mathbf{R}, \mathbf{R}^{-2}]_{\dot{\mathbf{R}}}\ .
\eqa

In practice, for a given a choice of $\mathbf{R}(t)$, the above equations must be solved in reverse order, \textit{i.e.}, first find $\mathbf{J}$ with Eq.~\eqref{Jeq}, then compute $\mathbf{A}$ using Eq.~\eqref{Aeq} and finally $\mathbf{M}$ with Eq.~\eqref{m_eq}.
Once $\mathbf{M}(t)$ is obtained in this fashion, the linear force term $\vec F$ is given by
\begin{equation}
    \label{F_eq}
    \vec{F} = m\left(\ddot{\vec{z}} + \mathbf{M}\vec{z} \right)\ ,
\end{equation}
in terms of the classical trajectory $\vec z(t)$~\cite{selw21}.

For any choice of $\vec z(t)$ and $\mathbf{R}(t)$, one can thus find a time-dependent Hamiltonian of the form given in Eq.~\eqref{Hamiltonian}, or equivalently Eq.~\eqref{Hamiltonian_2},
such that the eigenstates of the invariant are solutions of the Schr\"odinger equation.
Since a general choice of $\vec z(t)$ and $\mathbf{R}(t)$, however, does not ensure that the invariant commutes with $\hat H(t)$ at the start or end of the shuttling protocol,
it is not yet ensured that any eigenstate of the invariant is also an eigenstate of $\hat H(t)$ at the start or at the end of the shuttling protocol.
\begin{figure*}[t]
     \centering
\begin{subfigure}[b]{0.3\textwidth}
\centering
\includegraphics[width=\textwidth]{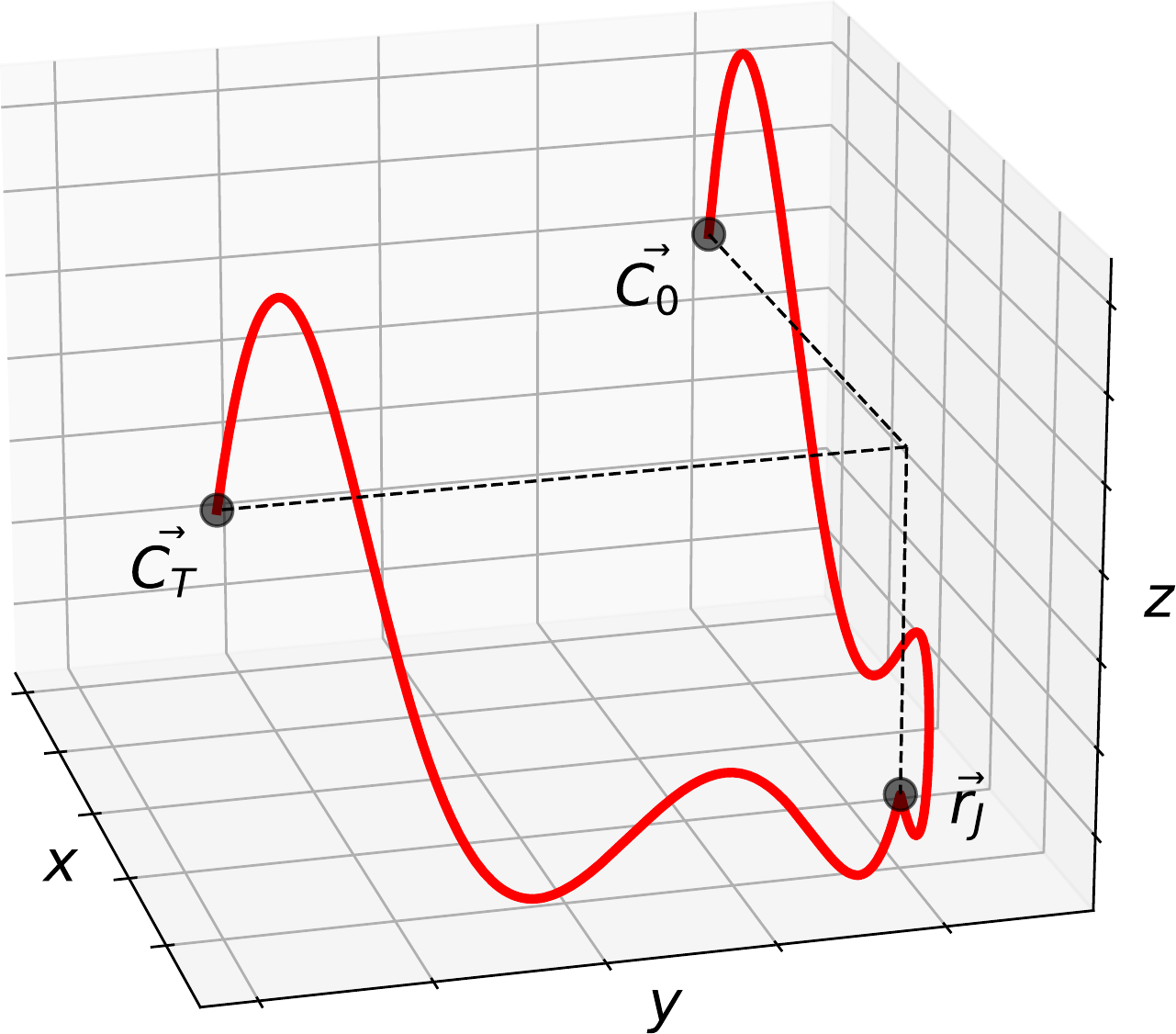}
\caption{}
    \label{fig:movement_ion}
\end{subfigure}
\hspace{0.05\textwidth}
\begin{subfigure}[b]{0.4\textwidth}
         \centering
\includegraphics[width=\textwidth]{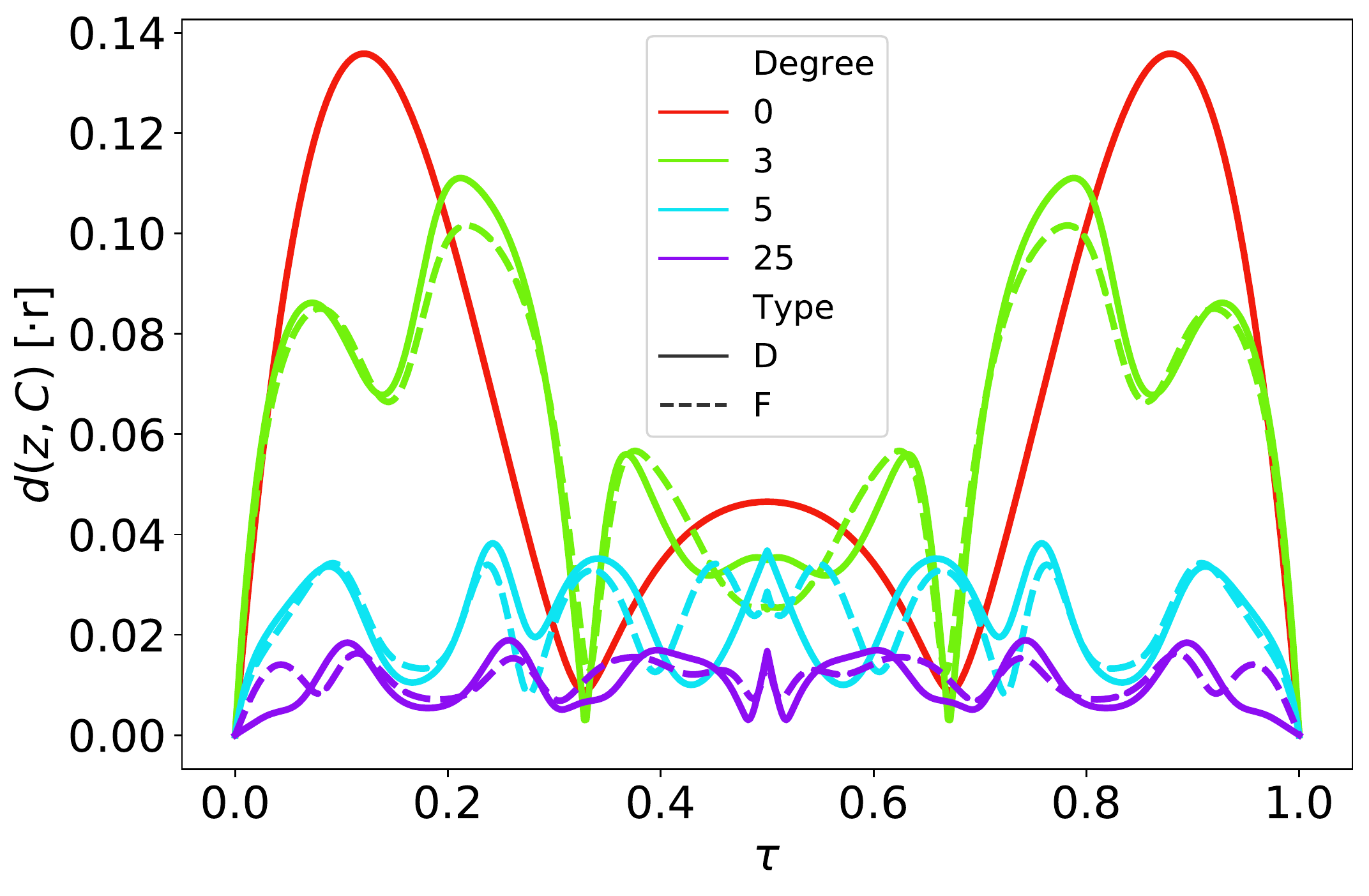}
\caption{}
    \label{fig:distances}
     \end{subfigure}
\caption{(a) Example of a corner trajectory (red solid line) from $\vec{C}_0$ to $\vec{C}_T$ restricted to pass through $\vec{r}_J$. Dashed black lines are drawn to facilitate the three-dimensional view of the problem. (b) Displacement as a function of normalized time ($\tau=t/T$). The solid lines correspond to the optimal solutions based on a diagonal form (D) of $\mathbf{R}$, while the dashed lines result from the optimisation of all entries of $\mathbf{R}$ (F). The color code distinguishes the number of control functions ($\it i.e.$, free parameters) considered in the optimisation.}
\label{fig:shuttling1}
\end{figure*}

Commutativity of the invariant and Hamiltonian at any instant $t$ is achived, if the set of relations
\begin{subequations}
\begin{alignat}{4}
\vec z(t)&=\vec{C}(t)\ ,\hspace{.5cm}&
\dot{\vec{z}}(t) = \ddot{\vec{z}}(t) = \vec{0}\ ,\label{eq:bcr}\\
\mathbf{R}(t) &= \mathbf{M}(t)^{-1/4}\ ,\hspace{.5cm}&
\dot{\mathbf{R}}(t)= \ddot{\mathbf{R}}(t)= \mathbb{0}_d\label{eq:bcR}
\end{alignat}
\label{eq:bc}
\end{subequations}
are satisfied~\cite{selw21}.
If the boundary conditions Eqs.~\eqref{eq:bc} are satisfied at $t=0$ and at $t=T$, with $\vec{C}(0)$, $\vec{C}(T)$, $\mathbf{M}(0)$ and $\mathbf{M}(T)$ determined by the initial and final Hamiltonian,
then the time-dependent ground-state of the invariant indeed defines a shuttling protocol in which an ion evolves from the ground-state of an initial Hamiltonian towards the ground-state of a final Hamiltonian.

\section{Optimal control}\label{intro_parametrisation}

The framework sketched above in Sec.~\ref{Invariant} enables the construction of time-dependent Hamiltonian resulting in ground-state to ground-state transfer.
Indeed, any choice of the time-dependent functions $\vec{z}(t)$ and $\mathbf{R}(t)$ results in a suitable Hamiltonian.
Since the shuttling protocol is thus anything but unique, one can aim at finding the protocol that is optimal in a sense to be specified. To this end, one would need to perform a variational analysis over $\vec{z}(t)$ and $\mathbf{R}(t)$ respecting the boundary conditions Eqs.~\eqref{eq:bc}.

For that purpose, it is convenient to use a set of functions $\{\vec{z}_i\}$ such that
$\vec{z}_0(t)$ satisfies the boundary conditions Eqs.~\eqref{eq:bcr}
and such that any function $\vec{z}_i(t)$ with $i>0$ satisfies the homogeneous boundary conditions, {\it i.e.}, Eqs.~\eqref{eq:bcr} with vanishing right-hand-sides.

The parametisation 
\be \label{manifold_z}
\vec{z}(t) = \vec{z}_0(t) + \sum_{i=1}^{N_{a}} a_{i}\vec{z}_i(t), 
\ee
then satisfies the boundary conditions for any set of $N_{a}$ expansion coefficients  $a_i$. 

A parametrisation of the matrix $\mathbf{R}(t)$ can be chosen in a similar way,
with one matrix $\mathbf{R}_0(t)$ satisfying the inhomogeneous boundary conditions, and a set of matrices $\mathbf{R}_i(t)$ satisfying the homogeneous version of Eqs.~\eqref{eq:bcR},
such that the parametrized matrix
\be\label{manifold_R}
\mathbf{R}(t)=\mathbf{R}_0(t) + \sum_{i=1}^{N_b} b_i\mathbf{R}_i(t) 
\ee
satisfies the boundary conditions for all values of the $N_b$ free parameters $b_i$.
The condition that $\mathbf{R}(t)$ be positive, however, requires some extra care. Demanding that all the matrices $\mathbf{R}_i(t)$ (including $i=0$) are positive, and that all the expansion coefficients $b_i$ are non-negative, does not give access to all positive matrices within the spanning set.
In all the following optimizations, there are thus no direct restrictions on the expansion coefficients $b_i$, but positivity of $\mathbf{R}(t)$ is assessed numerically, either in terms of sub-determinants (Sylvester's criterion, ~\cite{sylv85}), moments or numerically obtained eigenvalues.

\section{Results}\label{results}

The formalism introduced in the previous section is applicable to any control problem that requires the identification of ground-state to ground-state transport protocols.
In particular, it can be used to derive the time-dependent trapping potentials for ground-state-to-ground-state transport that is optimal according to a figure of merit to be specified.
The following section shows three examples of such optimized transport protocols.

\subsection{3D shuttling with reduced displacement from the potential center
}\label{corner_shut}

Realizing a fast, diabatic shuttling requires strong acceleration and deceleration at the beginning and the end of the protocol.
Such processes are most easily realised in terms of a large displacement of the ion from the center of the trapping potential.
Since in practice such large displacements imply that the ion traverses domains of substantial anharmonicity in the trapping potential, it is desirable to minimize the displacement.
The following discussion exemplifies the control problem of minimizing the maximum displacement of the ion while it moves around a corner in a potential landscape.

The initial and final position are denoted by the vectors  $\vec{C}_0=(0, r, h)$ and $\vec{C}_T=(r, 0, h)$. 
An additional boundary condition $\vec{r}_J(t=T/2)=(r, r, h')$ reflects the fact that realistic trapping potential do not always ensure dynamics with a constant value of the $z$-component of the ion trajectory (see Fig.~\ref{fig:movement_ion}).

The trapping potential along the ion trajectory is characterized in terms of a matrix $\mathbf{M}(t)$ following Eq.~\eqref{Hamiltonian_2}, that  satisfies the boundary conditions
\begin{eqnarray} \label{potenciales_3d}
\mathbf{M}_0 =
\begin{pmatrix}
\omega_t^2 & 0 & 0\\
0 & \omega_r^2 & 0 \\
0 & 0 & \omega_r^2
\end{pmatrix}
\mbox{ and }\
\mathbf{M}_T =
\begin{pmatrix}
\omega_r^2 & 0 & 0\\
0 & \omega_t^2 & 0 \\
0 & 0 & \omega_r^2
\end{pmatrix}
\end{eqnarray}
with and axial frequency $\omega_t$ that is substantially smaller than the radial frequency $\omega_r$,
so that the preferred direction of motion is initially along the $x$-axis and finally along the $y$-axis.
Since the displacement required for an acceleration or deceleration can be reduced to any desired value by increasing the confining potential along the axis of acceleration or deceleration, the following optimization includes the constraints that $\mathbf{M}_{xx}\le \min\left(\mathbf{M}_{yy},\mathbf{M}_{zz}\right) \le 2\omega_r$ for $t<T/2$ and $\mathbf{M}_{yy}\le \min\left(\mathbf{M}_{xx}, \mathbf{M}_{zz}\right) \le 2\omega_r$ for $t>T/2$.

With these boundary conditions and constraints, adiabatic shuttling yields close-to-perfect transport (with fidelities $F>99\%$, \cite{Note1}) in regimes where $T > 200/\omega_{t}$. Any transport with $T < 200/\omega_{t}$ is thus considered diabatic. The following results correspond to $T = 10/\omega_t$, {\it i.e.}, protocols an order of magnitude faster than in the adiabatic regime.

As described above in Sec.~\ref{intro_parametrisation}, both the classical trajectory of the ion $\vec{z}$ and the matrix $\mathbf{R}$ can be parameterized such that the boundary conditions are satisfied for any value of the expansion coefficients.
In order to highlight the ability of the present framework to find good solutions of a control problem in the presence of restrictions on achievable potentials,
the following discussion will compare two different parametrizations of $\mathbf{R}$.
In case $1$, the matrix $\mathbf{R}$ is taken to be diagonal, and only the time-dependent diagonal elements are subject to optimization.
{\bf The physical implication is that the principal axes of the trapping potential coincide with the $x$-, $y$- and $z$-axis.}
The potential can thus not be rotated, but its strength along the principal axes can be modulated.
In case $2$, each matrix element of $\mathbf{R}$ is subject to optimization, such that rotations of the principal axes are also possible.

Fig.~\ref{fig:distances} depicts the displacement, {\it i.e.} the distance between the ion and the center of the potential as a function of normalized time ($\tau=t/T$) for different solutions. The color code distinguishes the number of control functions in the optimization, while the line style differentiates solutions with $\mathbf{R}$ in diagonal (D) or full (F) form. The red curve depicts the solution without any optimization. For $\tau\lesssim 0.15$ the ion is being separated from the trap center, until a maximum separation of about $0.14 r$ is reached. Subsequently, the ion moves towards the trap center before it goes through another interval of separation. Due to the symmetry in trap geometry and time-dependence of the invariant, the trajectory is symmetric around the instant $\tau=1/2$. The large separation in the early and late part of the shuttling protocol is not unexpected, because the ion needs to be strongly accelerated and decelerated, which is easily achieved far away from the trap center where the trapping forces are strong.

\begin{figure}[t]
     \centering
\begin{subfigure}[b]{0.35\textwidth}
\centering
\includegraphics[width=\textwidth]{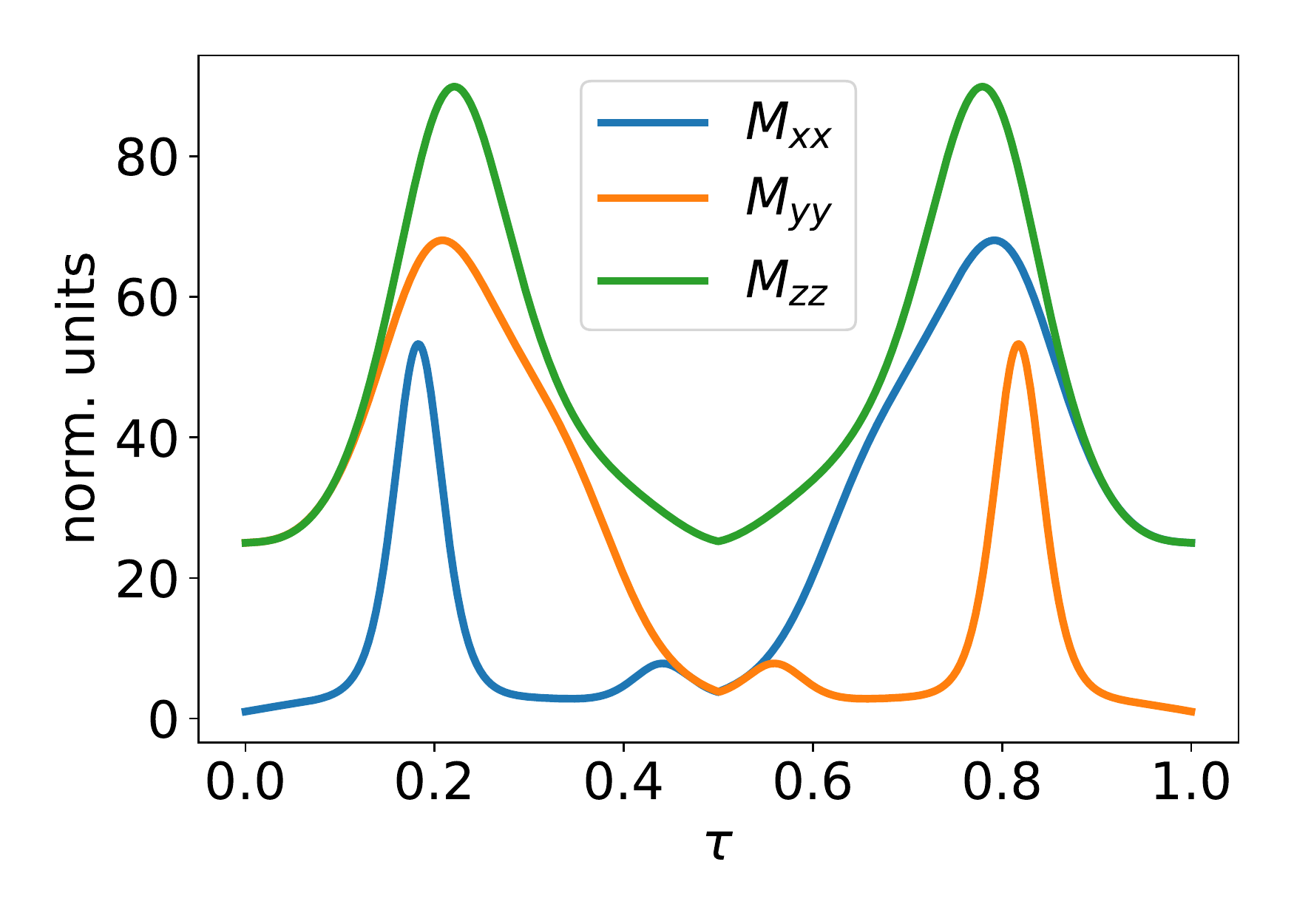}
\caption{}
    \label{fig:M_diag}
\end{subfigure}
\hspace{0.05\textwidth}
\begin{subfigure}[b]{0.35\textwidth}
         \centering
\includegraphics[width=\textwidth]{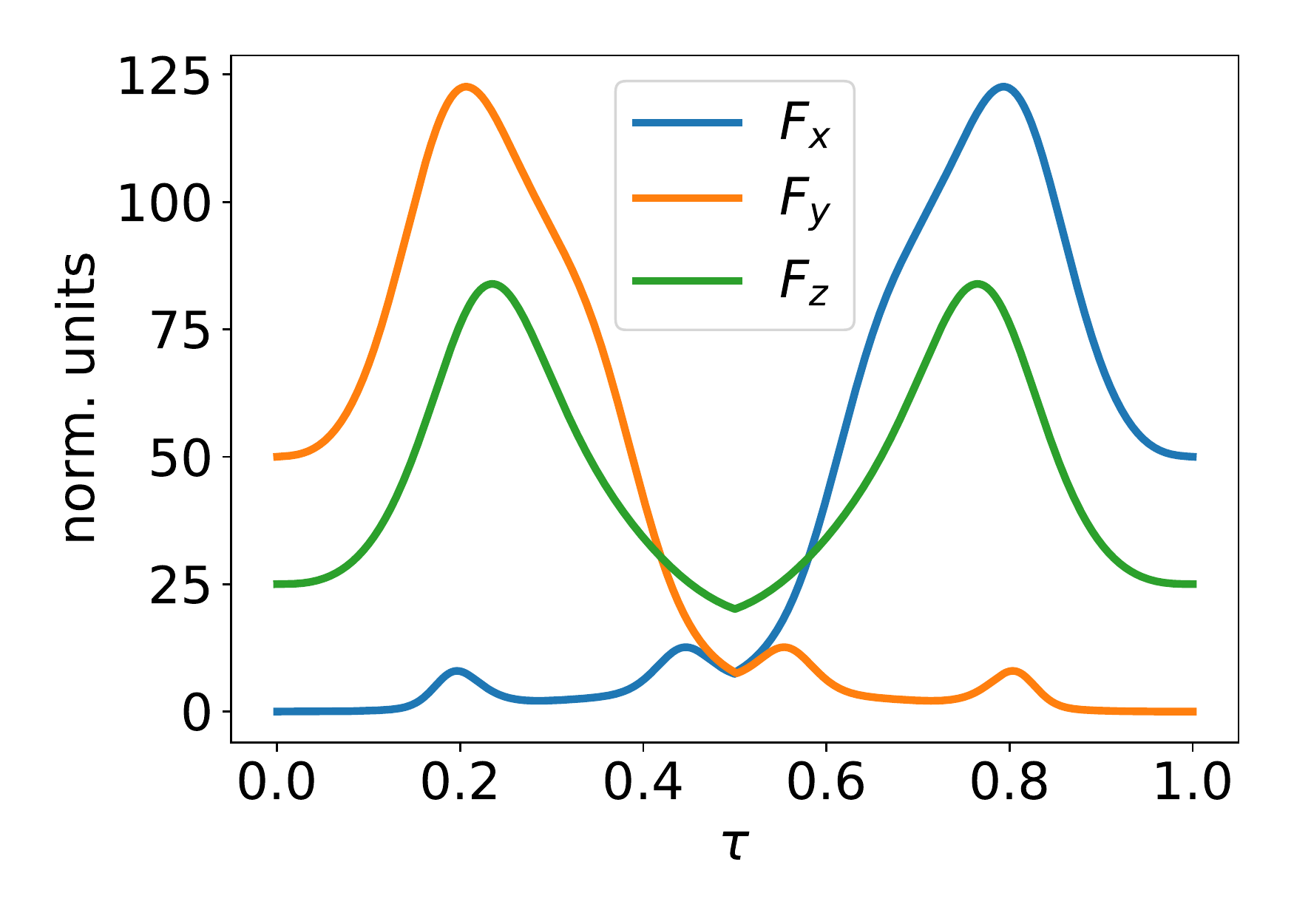}
\caption{}
    \label{fig:F_diag}
     \end{subfigure}
\caption{Optimal time evolution of (a) $M_{xx}$, $M_{yy}$, $M_{zz}$  and (b) $F_x, F_y, F_z$, in units of $\sqrt{\frac{\hbar}{m\omega_0}}$, for $\omega_r/\omega_t=5$ and $\omega_t T = 10$, based on $25$ control functions and $\mathbf{R}$ in diagonal form.}
\label{fig:R_diag}
\end{figure}

Achieving a similarly strong acceleration and deceleration without substantial displacement from the trap center requires a suitably designed time-dependence of the shuttling protocol. The green curves depict a numerically optimised protocol with $3$ control functions for each independent component of $\vec{z}$ and $\mathbf{R}$. Both forms of $\mathbf{R}$ lead to a reduction of the maximum displacement compared to the first solution. Nevertheless, rotating the trapping potentials (\textit{i.e.}, using the full $\mathbf{R}$ matrix, dashed line), slightly improves the protocol. Similarly, the optimization based on $5$ control functions (light blue curves) shows even greater improvement, as the displacement does not exceed $0.04r$. Likewise, releasing the major axes of the potential (dashed line) leads to a better solution. Further minimizing the displacement, however, requires a much more significant increase in the number of control functions. Thus, the optimisation of up to $25$ control functions (\textit{i.e.}, $225$ free parameters when the full $\mathbf{R}$ matrix is considered and $150$ otherwise) leads to a maximum displacement smaller than $0.02r$ (dashed purple line). The impact of employing rotating potentials is minor in this case. 

Unsurprisingly, at any given number of control functions (per matrix element), the maximum displacement obtained in an optimization with a general matrix $\mathbf{R}$ is slightly smaller than in the corresponding optimization with a diagonal matrix $\mathbf{R}$.
Crucially, however, optimizations with a diagonal matrix $\mathbf{R}$ can outperform optimizations with a general matrix $\mathbf{R}$ if the number of control functions in the former problem is only moderately larger than in the latter problem.
This means that restrictions in experimentally realizable rotating trapping potentials can be overcome in terms of additional temporal degree of freedom in the tuneable trapping parameters.

The discussion so far was focused on the trajectory on the ion, but a motional ground-state-to-ground-state protocol also requires the covariances of the Gaussian state to evolve towards their appropriate value.
Fig.~\ref{fig:R_diag} depicts the time-dependent components $M_{xx}$, $M_{yy}$ and $M_{zz}$ of the trapping potential in (a) and the linear force in (b).
In between the initial weak confinement along the $x$-axis and the final weak confinement along the $y$-axis,
there are two periods of time of enhanced confinement, reflecting the fact that strong confinement is necessary to rapidly accelerate or decelerate the ion. Fig.~\ref{fig:F_diag} shows that such an acceleration induces a force (see Eq.\eqref{F_eq}) that is particularly strong in the two directions subjected to change ($y,z$ in the first half and $x,z$ in the second). Merely enhancing the confinement and the force in order to ease control over the trajectory is, however not possible, since this would result in substantial undesired dynamics of the covariances.
Therefore, control of the covariances requires that strong confinement be applied only during some time windows, while weak confinement is necessary for the covariances to evolve in such a way that the ion ends up in the ground-state of the final Hamiltonian.

\subsection{Narrow wave-packets in weak confinement
}\label{min_uncertain}

The spatial uncertainty, {\it i.e.} the width of the wave-packet in diabatic dynamics is determined by the strength of the confining potential.
A weak potential implies a broad wavepacket that is susceptible to anharmonicties in the potential landscape.
In diabatic shuttling, however, it is possible that the wave-packet propagates through a domain of weak confinement without broadening to the natural width of its trapping potential.

Beyond the fact that a fast shuttling protocol leaves little time for the wave-packet to broaden, it is possible to initialise the wave-packet in a shape such that the natural dynamics makes it more narrow.
If this happens shortly before the wave-packet enters the domain of weak trapping potential, {\it i.e.} when it can still be controlled well by the stronger potential, the time-window in which the wave-packet remains narrow can be prolonged. 

In the following, this is exemplified with the problem of reducing the spatial width $\sigma_x$ of a wave-packet in a time-dependent potential trapping frequency $\omega(t)$.
The trapping frequency has to satisfy the boundary conditions $\omega(0)=\omega(T)=\omega_0$, and the wave-packet has to evolve from the ground-state of the initial Hamiltonian to the ground-state of the final Hamiltonian.
During the time intervals $[0,T/4]$ and $[3T/4, T]$, the trapping frequency can be modulated with a maximally allowed frequency $\sqrt{2}\omega_0$, but in the time-window $[T/4,3T/4]$, the maximally allowed trapping frequency is given by $\omega_0$.
The goal of the optimisation involves minimizing a weighted average of the maxima of $\sigma_x(t)$ and $\omega(t)$, \textit{i.e.}, the cost function $C = \alpha \cdot \max(\sigma_x(t)) + (1-\alpha) \cdot \max(\omega(t))$ at the central window $T/4<t<3T/4$ of the dynamics.  

\begin{figure}[t]
  \includegraphics[width=.4\textwidth]{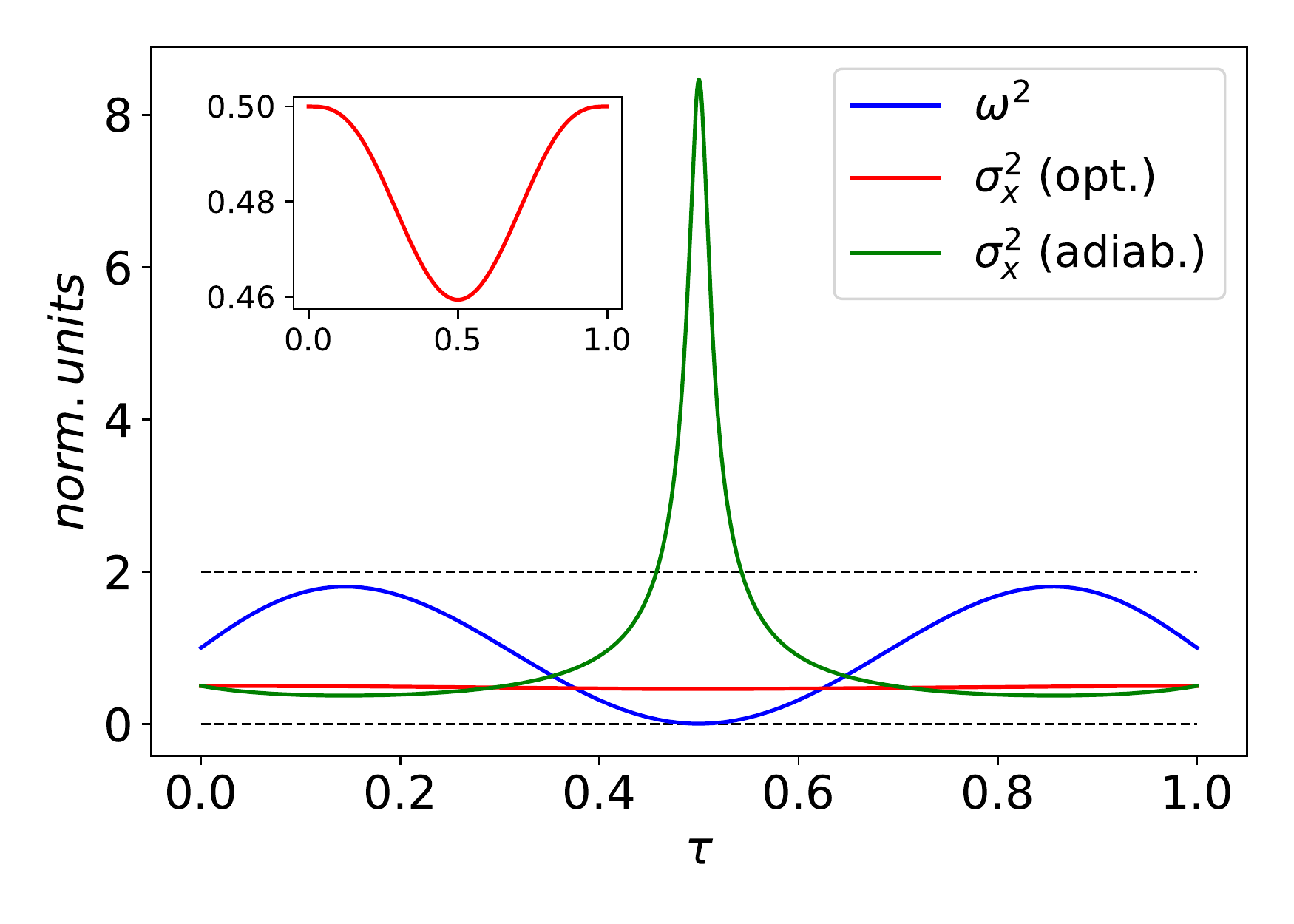}
    \caption{Squared trapping frequency $\omega^2$ and spatial uncertainty of the quantum state $\sigma_x^2$ as a function of normalized time. The red curve depicts the dynamics experienced by $\sigma_x^2$ when the evolution follows the optimal protocol, whereas the green curve corresponds to an adiabatic transport. The dashed black lines indicate upper and lower frequency limits and all quantities are expressed in units of $\sqrt{\frac{\hbar}{m\omega_0}}$, with $\omega_0 T = 1$ in this case.}
    \label{fig:weak_potential}
\end{figure}

\begin{figure*}[t]
     \centering
     \begin{subfigure}[b]{0.32\textwidth}
         \centering         \includegraphics[width=\textwidth]{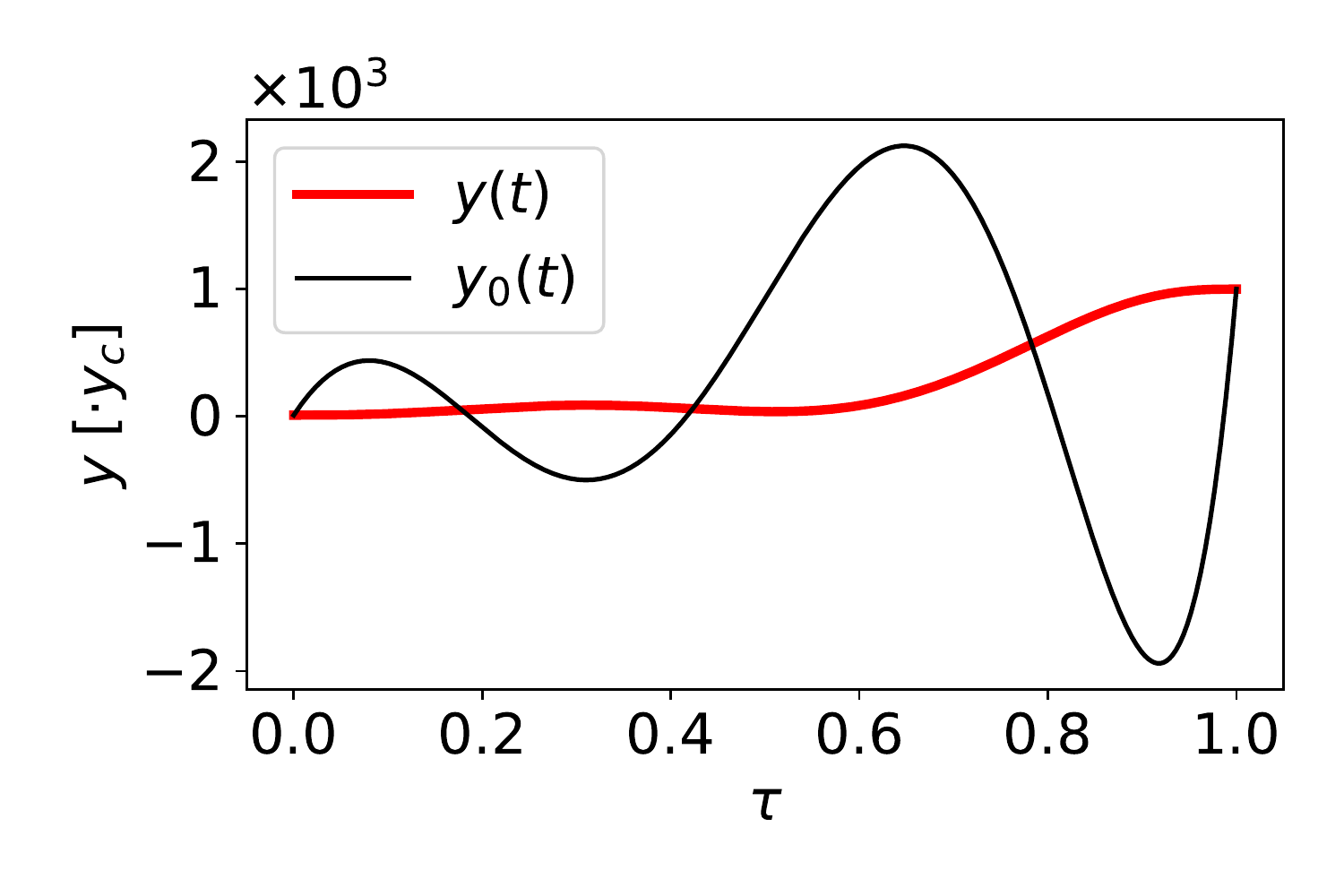}
         \caption{}
         \label{fig:Cxx}
     \end{subfigure}
     \hfill
     \begin{subfigure}[b]{0.32\textwidth}
         \centering
         \includegraphics[width=\textwidth]{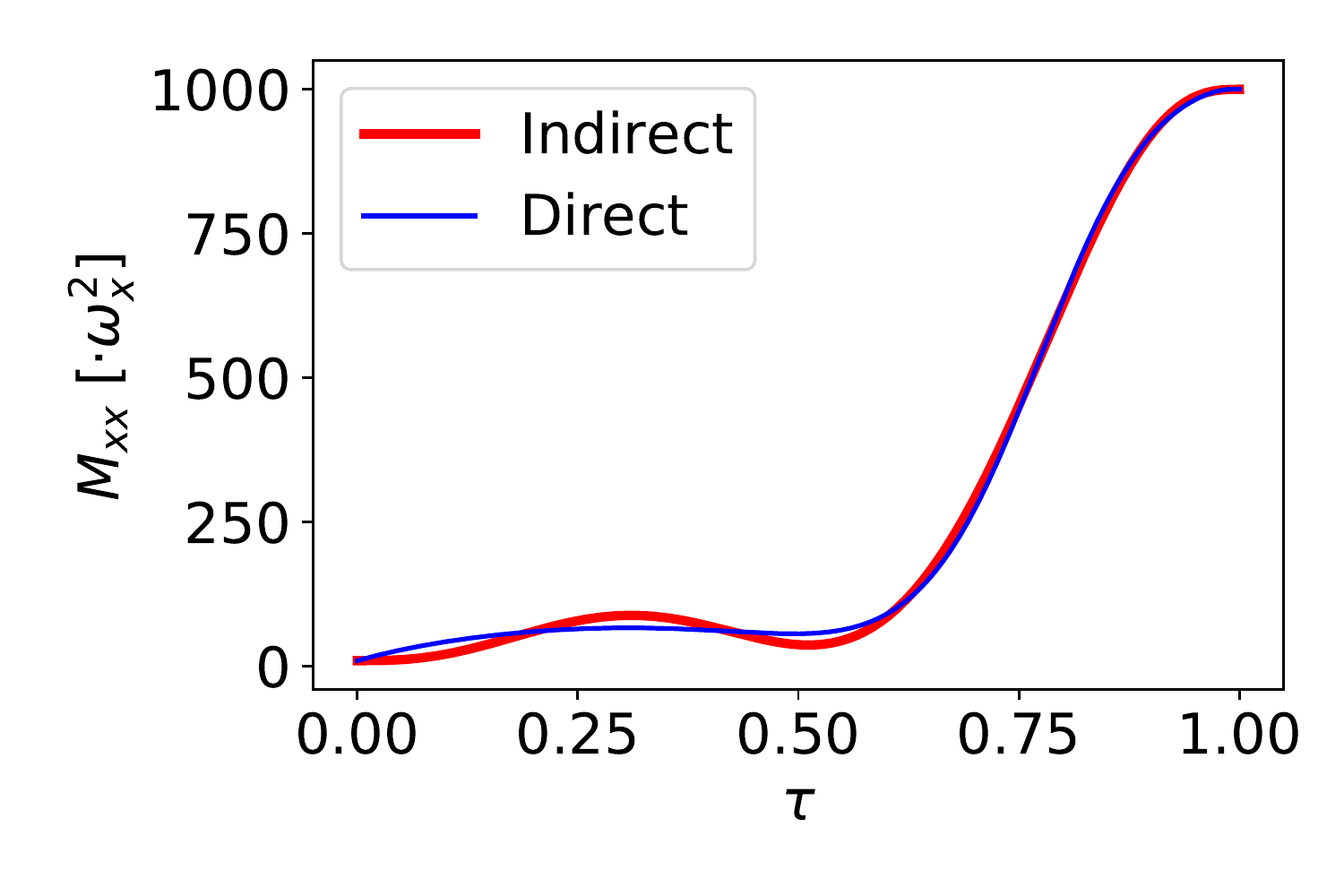}
         \caption{}
         \label{fig:Mxx}
     \end{subfigure}
     \hfill
     \begin{subfigure}[b]{0.32\textwidth}
         \centering
         \includegraphics[width=\textwidth]{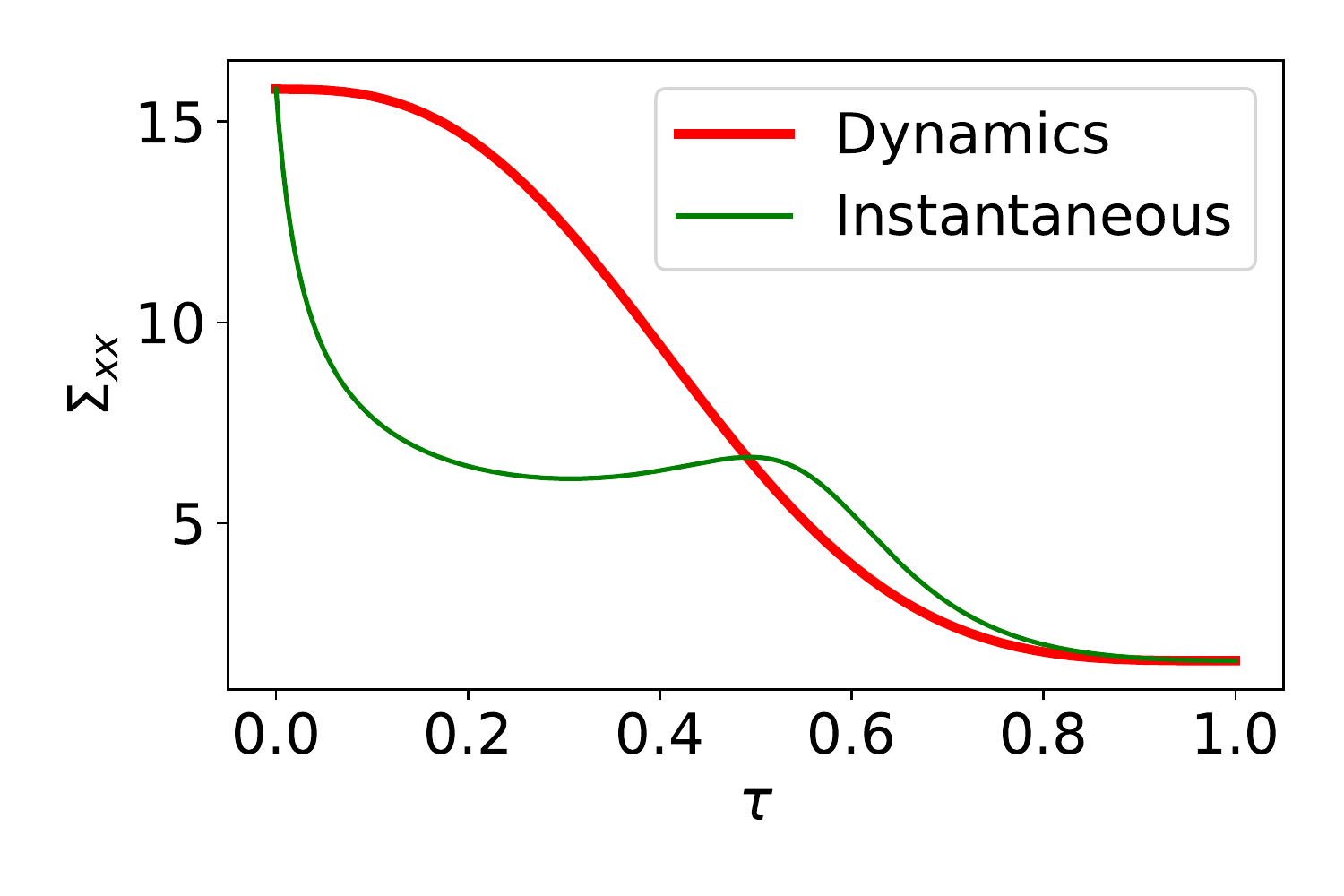}
         \caption{}
         \label{fig:Sxx}
     \end{subfigure}
        \caption{(a)  Trajectory of the ion along the $y$ axis as a function of time, following the optimal temporal evolution (red) and the corresponding trajectory of the trap center $y_0$ (black). (b) In red, curvature along the $x$-axis experienced by the ion as a function of time, achieved purely indirectly via the position-dependent curvature of the tapered trap. In blue, the curvature that guarantees perfect ground-state to ground-state transfer, obtained with direct dynamic control of the curvature independently of position.
        (c) Spread of the wave-packet in $x$ (in normalized units) as a function of time. The red line shows the dynamics resulting from following the trajectory $y(t)$ (with the appropriate velocity), while the green curve represents the instantaneous spread corresponding to the trajectory's curvature $M_{xx}$, ignoring the dynamics of the full trajectory.}
        \label{fig:taper}
\end{figure*}

The protocol shown in Fig.~\ref{fig:weak_potential} is based on the optimization of $25$ control functions. The left panel depicts the squared trapping frequency $\omega^2$ (in blue) as a function of normalized time. Its optimal evolution consists of increasing the strength of the potential to approximately twice the initial value and then gradually release the confinement so that in the central time-window ($T/4<t<3T/4$) the confinement is weak. The squared trapping frequency reaches a minimum near zero at $t=T/2$ and then exhibits a symmetric time evolution with respect to that time instance. The red curve in the figure represents the temporal evolution of the wave-packet spread in a potential evolving with the optimal $\omega^2(t)$. Not only does the uncertainty not increase throughout the protocol, but it even decreases slightly in the central region, when the potential is weak (see inset). Interestingly, the minimum uncertainty is obtained at $t=T/2$, coinciding with the point of minimum confinement. 

In weak confining potentials, wave-packets with narrow widths can only be obtained in the non-adiabatic regime. In fact, following the optimal time-evolution with a total protocol time 10 times longer,  $T'=10T$, radically different results are obtained. In this case, uncertainty evolves inversely proportional to confinement, as shown by the green curve in Fig.~\ref{fig:weak_potential}. In the first quarter of the protocol, it slightly decreases as a consequence of a greater confinement. As the potential strength is reduced, the wave packet spreads out more widely, reaching a maximum in the central window. Again, increasing the confining strength in the second half of the protocol results in a relocalization of the wave packet before it evolves to the final motional ground-state. 

\subsection{Shuttling based on null control over the curvature of the potential}\label{pseudo_matching}

The problem of shuttling around a corner discussed above in Sec.~\ref{corner_shut} highlights that restrictions in realizable trapping potential can be compensated by temporal degrees of freedom.
The ability to compensate lack of control over some degrees of freedom by controlling other degrees of freedom can be demonstrated in the following example, in which only linear force terms, but not confinement, are subject to controllable time-dependence.

Crucially, the dynamics of the controlled particle depends on the curvature along the classical trajectory taken by the particle.
If the actual trapping potential is anharmonic -- as it is the case in any realistic scenario -- then the particle can experience a position-dependent curvature, which can be controlled in a time-dependant way via the trajectory of the particle.
The choice of trajectory can therefore be used to compensate for restrictions on the tuneability of the trap confinement strength. 

This feature can be exemplified with the potential
\be \label{tapered_trap}
V(t)=\frac{1}{2}m\left(\omega_x^2 \frac{yx^2}{y_c}+\omega_y^2(y-y_0(t))^2\right)
\ee
of a tapered trap, in which the confinement in the $x$-direction varies along the $y$-axis. 
The trapping parameters $\omega_x$, $\omega_y$ and $y_c$ are time-independent,
and only the trap-center $y_0$ can be temporally modified.
The actual confinement in the $x$-direction is characterized by the frequency $\omega_x\sqrt{\frac{y(t)}{y_c}}$, and since the trajectory $y(t)$ can be controlled in terms of the trap center $y_0(t)$, is is possible to effectively modulate the confinement in the $x$-direction.

Because there is no direct control over the quadratic components of the trapping potential, the control framework with the present invariant does not provide a direct way to construct a protocol that guarantees ground-state-to-ground-state shuttling.
Applying the framework straight-forwardly results in an over-determination of the quadratic component of the potential, {\it i.e.} the matrix $\mathbf{M}$ in Eq.~\eqref{Hamiltonian}.
On the one hand, $\mathbf{M}$ is determined via Eq.~\eqref{m_eq} following the regular framework; but, on the other hand, it is also determined via curvatures of the trapping potential (Eq.~\eqref{tapered_trap}) along the ion trajectory $y(t)$.
The prescription for $\mathbf{M}(t)$ following Eq.~\eqref{m_eq} ensures ground-state-to-ground-state shuttling, but the prescription following Eq.~\eqref{tapered_trap} is what is experienced by the ion.
Without a means to ensure that those two prescriptions coincide, the requirement that the ion end up in the ground-state of the final potential is not automatically met by the basic framework. Nevertheless, the deviation between the two prescriptions can be numerically minimized over the degrees of freedom in selecting the time dependent trajectory $y(t)$, and a successful minimization will result in a ground-state-to-ground-state shuttling protocol.

In the following, this will be exemplified with the control problem of moving the trap center from its initial position $y_0(0)=y_i$ to its final position $y_0(T)=y_f$, such that the particle evolves from ground-state to ground-state of initial and final Hamiltonian.
Specific parameter values used in the example are $y_i=10y_c$, $y_f=1000y_c$ and $\omega_y=10\omega_x$.
With those parameter values, adiabatic transport is considered to occur when  $T>70/\omega_y$ (for which $F>99\%$, or, equivalently, a final motional heating $\bar n<10^{-2}$). The results shown here correspond to $T=3/\omega_y$ (regime in which extending the adiabatic protocol leads to $\bar n = \mathcal{O} (10^{3})$).  

Fig.~\ref{fig:taper}(a) depicts the solution obtained for the trap center $y_0(t)$ in black.
It exhibits oscillations far outside the interval through which the ion is meant to be shuttled.
The ion trajectory (depicted in red), however, remains within this interval, and the motion of the ion is comparatively slow in the first $60\%$ of the time-window.
Most of the transport occurs within a short time-window (about $20\%$ of the full duration), and in the last $10\%$ of the time-window the ion is, again, moving very slowly.

As a result of this dynamics, the confinement along the $x$-direction remains rather weak in the first half of the shuttling protocol, and it increases towards its final value only later, as depicted in Fig.~\ref{fig:taper}(b) in red.
The time-dependent confinement that ensures perfect ground-state-to-ground-state transfer (\textit{i.e.}, derived from the invariant framework) is depicted in blue. 
The deviation between the two is fairly minor (in fact, the infidelity of the protocol is below $0.02\%$, \textit{i.e.}, the final motional excitation $\bar n < 10^{-4}$), meaning that the dynamics experienced by the ion will be very similar to that which guarantees ground-state-to-ground-state transfer in the case of direct control of the curvature of the potential. 

The close-to-perfect fidelity of the shuttling protocol also implies that the covariances of the ion evolve towards the values corresponding to the ground-state.
This can also be seen in Fig.~\ref{fig:taper}(c).
The green curve in Fig.~\ref{fig:taper}(c) depicts
the {\it instantaneous} spatial covariance, {\it i.e.} the covariance corresponding to adiabatic transport along the actual trajectory of the ion (as depicted in red in Fig.~\ref{fig:taper}(a)). The red curve depicts the dynamics of $\Sigma_{xx}$ following the same trajectory, but with the actual velocity.
Despite the strong acceleration of the ion (corresponding to the high amplitudes of the trap center in Fig.~\ref{fig:taper}(a)),
and the periods of slow dynamics of the ion, that results in the plateau of the instantaneous covariance (green in Fig.~\ref{fig:taper}c)), the actual dynamics of $\Sigma_{xx}$ is rather unspectacular, {\it i.e.} the width of wave-packet simply shrinks monotonically and adopts its desired final value at the end of the shuttling protocol.
The covariance thus behaves exactly the way that one would have designed it in an adiabatic protocol, even though the protocol is far outside the regime of adiabaticity.

\section{Conclusions}

Ground-state-to-ground-state shuttling remains one of the crucial challenges in the realization of scalable quantum information processing of trapped ions.
Invariant-based quantum control is a promising route towards finding shuttling protocols that facilitate the suppression on undesired motional excitations.
Given the multiple experimental imperfections that can contribute to reduced quality of shuttling operations, the ability to optimize shuttling protocols without compromising the goal to have an ion end up in its ground state can become a central step towards a practical technology.

The exemplary tasks of reducing the displacement from the center of the trapping potential and the spatial width of a wave-packet without strong confining potential can give a flavor of what can be achieved with suitable temporal shaping of trapping potentials.

While the framework of invariants can ensure that the resultant Hamiltonians satisfy physically motivated constraints, such as the decomposition into a pre-defined kinetic energy term and a time-dependent potential term, any practical implementation typically requires compliance with additional constraints resultant {\it e.g.} from the geometry of the trap electrodes. 
Given the notorious difficulty to construct an invariant that is consistent with a Hamiltonian of a specified set of properties, it is hopeless to expect that suitable invariants for given trap geometries can be found.
The ability to minimize the deviations between solutions of the general invariant framework and solutions that can be obtained in the presence of experimental constraints, however, enables the incorporation of such constraints in practice.


\appendix
\section{Parametrisation of $\vec{z}(t)$ and $\mathbf{R}(t)$}\label{appendix:a}

As explained in Section $\ref{intro_parametrisation}$, a quantum invariant compatible with a quadratic Hamiltonian is determined by the choice of $\vec{z}(t)$ and $\mathbf{R}(t)$, subject to satisfy a set of boundary conditions. Any solution for $\vec{z}(t)$ can be parametrised in terms of a set function $\vec{z}_{i}(t)$ (with $i\geq 0$):
\be 
\vec{z}(t)= \vec{z}_0(t) + \sum_{i=1}^{N_a} a_i\vec{z}_i(t)
\ee
where $\vec{z}_0(t)$ satisfies the inhomogeneous boundary conditions (Eqs.~\eqref{eq:bcr}) and $\vec{z}_{i>0}(t)$ fulfil the corresponding homogeneous equations.  Thus, $\vec{z}_0(t)$ can be chosen to be the lowest-degree polynomial satisfying the associated boundary equations in each component, $\it{i.e.}$ parametrizing each component of $\vec{z}_0(t)$ with the following function:
\begin{dmath} \label{parametrise_f}
    f(\tau) = C_{I} + 10(C_F - C_I)\tau^3 - 15(C_F - C_I)\tau^4 + 6(C_F - C_I)\tau^5 
\end{dmath}
where $C_I, C_F$ refer to the initial and final values respectively and $\tau\equiv \frac{t}{T}$ is the normalized time. If extra constraints are imposed beyond the boundary conditions, polynomials of higher degree must be considered. For example, in the case of constraining $f(\tau=1/2)=C_J$, the function that fulfills all the constraints is
\begin{dmath}  \label{parametrise_f_tilde}
    \tilde f(\tau) = C_I + 2(-11C_F - 21C_I + 32C_J)\tau^3 + (81C_F + 111C_I - 192C_J)\tau^4 + 2(-45C_F - 51C_I + 96C_J)\tau^5 + 32(C_F + C_I - 2 C_J)\tau^6 .
\end{dmath}
The rest of functions $\vec{z}_{i>0}$ satisfy the homogeneous boundary conditions. A similar polynomial expansion may be used for each of the components of $\vec{z}_{i>0}$, but the following parametrisation is more suitable for problems with fast dynamics:

\begin{dmath} \label{parametrise_g}
    g(\tau) =  \sin \left(\pi(1 + 2\omega_2)\tau \right) - \frac{1 + 2\omega_2}{1 + 2\omega_1}\sin(\pi(1 + 2\omega_1)\tau) ,
\end{dmath}
where the coefficients $\omega_1, \omega_2$ are free provided that $\omega_1, \omega_2 \in \mathbb{Z}$. Therefore, $g(\tau)$ defines a set of vectors $\vec{z}_{i>0}$ that result in a valid solution for $\vec{z}(t)$ independently of the expansion coefficients $a_i$. 

As in the case of $\vec{z}_0$, it may also necessary to impose constraints on the functions $g(\tau)$ for some particular control problems. Of particular interest is to impose that $g(\tau=1/2)=0$. In that case, a set of functions that satisfy simultaneously the homogeneous boundary conditions and the added constraint is

\begin{dmath} \label{parametrise_g_tilde}
    \tilde g(\tau) = 
    \frac{\omega_2 S_3-\omega_3S_2}{\omega_1S_2-\omega_2S_1}\sin(\pi\omega_1\tau) + \frac{-\omega_1S_3+\omega_3S_1}{\omega_1S_2-\omega_2S_1}\sin(\pi\omega_2\tau) +  \sin(\pi\omega_3\tau), 
\end{dmath}
where we introduced the notation $S_i \equiv \sin(\pi\omega_i/2)$ and the frequencies $\omega_1, \omega_2, \omega_3 \in \mathbb{Z}$ subject to $\omega_1 \neq \omega_2$. 

Similarly, the positive semi-definite matrix $\mathbf{R}(t)$ can be parametrised as

\be
\mathbf{R}(t) = \mathbf{R}_0(t) + \sum_{i=1}^{N_b} b_i\mathbf{R}_i(t),
\ee

where $ \mathbf{R}_0(t)$ is a $d\times d$ diagonal matrix that fulfils the non-homogeneous boundary conditions and can be conveniently expressed as
\begin{eqnarray}
\mathbf{R}_{0}(\tau) = 
    \begin{pmatrix}
    f_1(t) & 0 & 0\\
    0 & f_2(t) & 0 \\
    \vdots & \ddots & \vdots \\
    0 & \cdots & f_d(t) 
    \end{pmatrix}, 
\end{eqnarray} 
where $f_i(t)$ are polynomials of $p$-degree satisfying that $f_i(t)>0, \forall i, t$. The boundary conditions are imposed as $f_i(0)=M_{I, ii}^{-1/4}, f_i(T)=M_{F, ii}^{-1/4}, \dot{f}_i(0)=\ddot{f}_i(0)=\dot{f}_i(T)=\ddot{f}_i(T)=0$. These conditions are fulfilled by the control function defined in Eq.~\eqref{parametrise_f}. Note that the non-negativity is also guaranteed because $C_I,C_F>0$ and the control function in Eq.~\eqref{parametrise_f} has no minimum for $\tau \in (0,1)$. 

On the other hand, the set of matrices $\mathbf{R}_{i}(t)$ ($i>0$) can be chosen following two different alternatives. The first option is to assume each $\mathbf{R}_{i}(t)$ as a matrix with functions in each entry that satisfy the homogeneous boundary conditions. Then, $\mathbf{R}(t)>0$ is numerically assured at every instant of time either in terms of the eigenvalues, the Sylvester's criterion or the calculation of moments.  

Alternatively, each matrix $\mathbf{R}_{i}(t)$ ($i>0$) can be decomposed in terms of lower-triangular matrices $\mathbf{L}_{i}$ such that $\mathbf{R}_{i}(t)=\mathbf{L}_{i}\mathbf{L}_i^{\dagger}$ satisfy the homogeneous boundary conditions and ensure that $\mathbf{R}(t)>0$. As
\begin{eqnarray} 
\mathbf{L}_{i}(\tau) = 
    \begin{pmatrix}
    g_{11}(t) & 0 & 0\\
    g_{21}(t) & g_{22}(t) & 0 \\
    \vdots & \ddots & \vdots \\
    g_{d1} & \cdots & g_{dd}(t) 
    \end{pmatrix}, 
\end{eqnarray} 
it is sufficient to require all the functions $g_{ij}(t)$ to satisfy the homogeneous boundary conditions to guarantee that $\mathbf{L}_i\mathbf{L}_i^{\dagger}$ does so. In addition, in order for $\mathbf{L}_i\mathbf{L}_i^{\dagger}$ to be positive semi-definite, the diagonal must be non-negative throughout the time domain. Both constraints are fulfilled if the off-diagonal functions are chosen from Eq.~\eqref{parametrise_g} and the diagonal is parametrised in term of the non-negative functions $h(\tau) \equiv g(\tau)^2$, where $g(\tau)$ is defined in Eq.~\eqref{parametrise_g}. Therefore, restricting the expansion coefficients $b_i$ to be positive guarantees that $\mathbf{R}(t)$ is positive semi-definite in the whole time domain. Note that, although this method guarantees the positivity of $\mathbf{R}(t)$ at any time, also limits the spectrum of valid solutions.  

\bibliography{biblio}

\end{document}